\documentclass{article}

\usepackage{arxiv}

\usepackage[utf8]{inputenc} 
\usepackage[T1]{fontenc}    
\usepackage{hyperref}       
\usepackage{url}            
\usepackage{booktabs}       
\usepackage{amsfonts}       
\usepackage{nicefrac}       
\usepackage{microtype}      
\usepackage{lipsum}
\usepackage{cite}
\usepackage[pdftex]{graphicx} 
\usepackage{amsmath}
\usepackage{graphicx}
\usepackage[section]{placeins}
\usepackage{multirow}
\usepackage{multicol}

\usepackage{breakurl}

\title{Will AI Make Cyber Swords or Shields \\
\large A few mathematical models of technological progress}

\author{
  Andrew J. Lohn and Krystal Alex Jackson\\
  Center for Security and Emerging Technology, Georgetown University \\
  \texttt{drew.lohn@georgetown.edu} \\
}

\begin{document}
\maketitle


\begin{abstract}
    We aim to demonstrate the value of mathematical models for policy debates about technological progress in cybersecurity by considering phishing, vulnerability discovery, and the dynamics between patching and exploitation. We then adjust the inputs to those mathematical models to match some possible advances in their underlying technology. We find that AI's impact on phishing may be overestimated but could lead to more attacks going undetected. Advances in vulnerability discovery have the potential to help attackers more than defenders. And automation that writes exploits is more useful to attackers than automation that writes patches, although advances that help deploy patches faster have the potential to be more impactful than either.
\end{abstract}

\section{Introduction}
\label{sec:introduction}
Predicting the impact of advances in technology may be a fool's errand but it is a necessary one nonetheless to help try to guide research and funding toward efforts that benefit defense more than offense. For this paper, we try to mathematically model the impact of further advancement in several critical aspects of cybersecurity. Perhaps more importantly than any of the forewarnings or funding recommendations we come to, this approach strives to sharpen debates about AI's impact on cybersecurity. This is the companion paper for a separate report, published by CSET and titled, "Will AI Make Cyber Swords or Shields," illustrating the value of rigor in policy discussions about technological advancement. There is too much uncertainty to believe that the math gives precise projections, but it forces us to be precise in our assumptions. Reasonable people may disagree with the range of values we choose as inputs or even the models we use. We welcome those disagreements and hope they advance our collective understanding of how AI may change the future of cybersecurity. 

Following this introduction, we proceed with separate analysis from three areas of cybersecurity: 1) phishing, 2) vulnerability discovery, then 3) the dynamics between patching and exploitation. 

List of variables:

\begin{tabular}{c c l}
   \textbf{Symbol}  & \textbf{Section Name} & \textbf{Description}\\
    $P_{infection}$ & Phishing & Probability of at least one successful phish in the victim organization\\
    $P_{click}$ & Phishing & Probability any one individual will be successfully phished \\
    $P_{no\_alert}$ & Phishing & Probability that a phishing campaign is not reported\\
    $P_{human\_alert}$ & Phishing & Probability that at least one human reports the campaign\\
    $P_{machine\_alert}$ & Phishing & Probability that a computer identifies at least one message as malicious\\
    $P_{undetected\_infection}$ & Phishing & Probability of at least one successful phish with none reported\\
    $R$ & Vulnerability Discovery & Rate of vulnerability discovery (discoveries per weeks)\\
    $S$ & Vulnerability Discovery & Number of unique vulnerabilities discovered\\
    $C_{time}$ & Vulnerability Discovery & Expected number of vulnerabilities discovered in the first week\\
    $C_{attempts}$ & Vulnerability Discovery & Expected number of vulnerabilities discovered in the first attempt\\
    $t$ & Vulnerability Discovery & Time (indicating the time passed while searching for vulnerabilities)\\
    $n$ & Vulnerability Discovery & Count for number of attempts\\
    $T$ & Vulnerability Discovery & Time (as bounds evaluating vulnerability discoveries within a time interval)\\
    $N$ & Vulnerability Discovery & Count of attempts (as bounds for evaluating number of vulnerability discoveries)\\
    $\eta$ & Vulnerability Discovery & Rate of vulnerability discovery attempts\\
    $\alpha$ & Vulnerability Discovery &  Shape factor for power law\\
    $P_{patch\_developed}$ & Patching vs Exploitation & Probability a vulnerability has a corresponding patch\\
    $k$ & Patching vs Exploitation & Shape parameter (degree patches more prioritized early)\\
    $\lambda$ & Patching vs Exploitation & Scale parameter (adjusts the characteristic delay for patches)\\
    $P_{patch\_deployed}$ & Patching vs Exploitation & Fraction of systems that have adopted an existing patch\\
    $\beta$ & Patching vs Exploitation & Time constant for exponential decay\\
    $P_{patch\_delay}$ & Patching vs Exploitation & Probability a system is patched for a specific vulnerability\\
    $P_{exploit\_development}$ & Patching vs Exploitation & Probability that an exploit exists for a specific vulnerability\\
    $A$ & Patching vs Exploitation & Linear coefficient for exploit development\\
    $a$ & Patching vs Exploitation & Power law exponent for exploit development\\
    $b$ & Patching vs Exploitation & Exponential decay rate for exploit development\\
    $\tau$ & Patching vs Exploitation & dummy variable \\
    $f$ & Patching vs Exploitation & dummy function\\
\end{tabular}

\section{Phishing}
\label{sec:phishing}
With AI's improving ability to write convincing messages, phishing is a natural area expect changes but it is already one of the most common ways to gain access to a system or organization.\cite{GPT3_Phishing} Given a large enough phishing campaign, attackers are almost guaranteed to win a foothold somewhere. That is true despite the fall in click rates from about 25\% to only about 3\% as education programs and general digital literacy have improved.\cite{2021_dbir} The falling click rates may not keep dedicated attackers out but they do provide some defensive advantage in that more messages must be sent to guarantee success. That increases the odds of alerting the victim's defense teams who can act quickly to block the messages and isolate or clean infected systems. This may partly explain why the odds of a phishing campaign being reported at least once have been rising from about 20\% in 2016 to almost 40\% in 2019.\cite{2020_dbir}  

We calculate the probability of a phishing campaign successfully gaining a foothold while avoiding being detected as the probability of at least one message being clicked times one minus the probability of at least one message being reported by either a human or a machine.

\begin{equation}
    P_{infection} = 1 - (1-P_{click})^{N}
\end{equation}
\begin{equation}
    P_{no\_alert} = (1 - (P_{human\_alert} + P_{machine\_alert} - P_{human\_alert} * P_{machine\_alert}))^{N}
\end{equation}

\begin{equation}
    P_{undetected\_infection} = P_{infection} * P_{no\_alert}
\end{equation}

There are variations of this model that could be considered. For one, our approach treats each message as independent. That is because we are imagining that each message as uniquely crafted by the AI so that correlations between them are limited and we imagine that detection is based on the content of the message. Alternatively, it may be reasonable to assume that the messages are correlated so that each new message changes the overall probability by less. Or perhaps the machines use metadata rather than content so the additional messages add little. We also assume that the probability of alert for humans and machines are not correlated. That simplifies the model but could be easily changed. Additionally, we assume that once one message is reported, defenders can use that information to find all the others which may not be true. These variations of our model are all interesting to consider and we encourage others to do so. For now, we explore the model with independent messages where identifying one message enables flagging them all.

\begin{figure}[h]
    \centering
    \includegraphics[width=8cm]{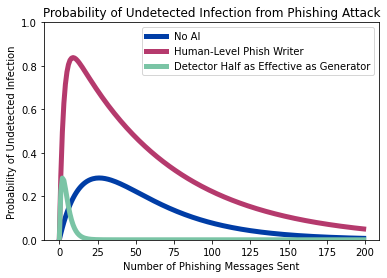}
    \caption{Phishing campaigns are more likely to go undetected with improved writing but fewer messages are required. Improvements in automated detection can bring down both success rates and the optimal number of messages to send.}
    \label{fig:phishing}
\end{figure}

As an approximate contemporary human baseline to compare against, we set $P_{click}=0.03$, $P_{human\_alert}=0.015$, and $P_{machine\_alert}=0.01$. In practice, these parameters vary across different phishing campaigns and victim organizations. Our focus is not in setting the parameters precisely but in understanding how changes in those parameters affect security outcomes. Along those lines, Figure \ref{fig:phishing} shows this baseline along with two AI-enabled projections: 1) an AI that can write phishing messages as effectively as humans can spear-phish ($P_{click}=0.3$ and $P_{human\_alert}=0.005$), and 2) where the machine-based detector improves along with the writer ($P_{machine\_alert}=0.25$).\cite{spearphish_susceptibility, tailored_phishing} We could alternatively imagine an AI outperforming humans in phishing campaigns because of its ability to draw on more information about its victims from across the digital domain. We may also be overly optimistic by setting the machine alert rate to 25\% given how much machines have struggled with attributing AI-generated text, although there is potential for improvements in using non-linguistic cues such contents of the headers. Regardless, these three example curves help to illustrate the range of phishing effectiveness.

\begin{table}[h]
    \centering
    \begin{tabular}{|c|c|c|c|}
        \hline
         & $P_{click}$ & $P_{human\_alert}$ & $P_{machine\_alert}$ \\ \hline
        No AI & 3\% & 1.5\% & 1\% \\ \hline
        Human-Level Phish Writer & 30\% & 0.5\% & 1\% \\ \hline
        Detector Nearly as Effective as Generator & 30\% & 0.5\% & 25\% \\ \hline
    \end{tabular}
    \caption{Parameter values for Figure \ref{fig:phishing}}
    \label{tab:phishing_parameters}
\end{table}

All three curves in \ref{fig:phishing} rise to a peak then fall with a long tail as the number of sent messages increases. That is because multiple messages are needed to increase the odds of at least one message being acted on by the victim, but too many messages increases the odds of the defenses being alerted. For the no-AI baseline, that peak is at 26 messages with a 28\% chance of an undetected intrusion. An AI that writes phishing messages as well as humans can spear-phish raises those odds to about 84\% with only nine messages. The introduction of 25\% effective automated phishing detection technology reduces the undetected intrusion rate back to 28\% with a peak at two sent messages. 

The current base rate of 28\% is worryingly high. A jump to 84\% introduces a new degree of risk in that phishing campaigns, which are already regularly successful might be very unlikely to detect. It takes optimistic projections of the defensive technology to return that rate to 28\%, although since those automate approaches peak with only a handful of messages, the value of automation is somewhat reduced because humans can write a handful of messages (consider defensive advancement without attacker progress). That is at least if a single organization is targeted. The automation could write a handful of high quality messages for a large number of organizations, particularly given how much personal data is available to shape those messages. Organizations that would otherwise have a low profile may now be attacked as if they were being individually targeted.

\section{Vulnerability Discovery}
\label{sec:vulndisc}
Once an attacker gains their foothold, or as an initial step instead of phishing, they can exploit vulnerabilities in a user or organization's software. These vulnerabilities are bugs that allow nefarious actors to do malicious things that the designers did not intend. Finding and managing these vulnerabilities is one of the central tasks of cybersecurity. Attackers are at an advantage when they find those vulnerabilities first.

When new software is released, some of its vulnerabilities are easier to find and others take more time to uncover. The decrease in discovery rate follows a power law distribution whose fat-tail means that although the discovery rate decreases over time, there are some vulnerabilities that will be discovered after very long times.\cite{vuln_disc_power_law, vuln_disc_is_power_law} The power law applies for black-box mutational fuzzing, which is a simple computer-based automated search technique.\cite{bb_mut_fuzz} It also applies for bug bounty programs where a collection of humans submit vulnerabilities they find to defenders for possible payment.\cite{vuln_disc_bounties} The parameters ($\alpha$) for each scenario are different but the mathematical model is the one shown in Eqn \ref{eqn:power_law}. It can be thought of in terms of either time or number of attempts by assuming a constant rate of attempts ($\eta$) that can convert between a total amount of time ($T$) or a total number of attempts ($N$) in Eqn \ref{eqn:power_law_N-T}.

\begin{equation}
    \label{eqn:power_law}
    R = C_{time}t^{-\alpha} = C_{attempts}n^{-\alpha}
\end{equation}
\begin{equation}
    \label{eqn:power_law_N-T}
    N=\eta T
\end{equation}

In this equation, the rate of discoveries ($R$) is driven by two parameters: $C$ can be simply thought of as an initial rate, and $\alpha$ reflects how much more difficult the next vulnerability is to find than the previous one. Both of these parameters depend on the software being tested and the capability of the tester. A large $C$ can indicated that the software has many vulnerabilities or that the testers that are good at finding vulnerabilities. A large $\alpha$ indicates that some vulnerabilities are much easier to find than others or that the tester is good at finding some types of vulnerabilities but not at finding different ones. A small $\alpha$ indicates a creative tester that is adept at finding new vulnerabilities after easier-to-find ones have already been uncovered. Based on experiments, the black-box mutational fuzzing approach had $\alpha$ in the range of 2 to 4 for various pieces of software, and the set of humans who submitted to a bug bounty program had $\alpha$ around 0.4 as a collective. 

The papers cited above do not provide values for $C$ but they can be estimated from the data in the papers. The number of vulnerabilities they found is over a specified search duration. The equation for the expected number of vulnerabilities, or unique species of vulnerabilities ($S$), over a search period we need to take the integral of Eqn \ref{eqn:power_law} as shown in Eqns \ref{eqn:vuln_disc_integral_N} and \ref{eqn:vuln_disc_integral_T}.

\begin{equation}
    \label{eqn:vuln_disc_integral_N}
    S = \int_{N_1}^{N_2} C_{attempts}n^{-\alpha} = \frac{C_attempts}{1-\alpha}[N_{2}^{1-\alpha}-N_{1}^{1-\alpha}]
\end{equation}
\begin{equation}
    \label{eqn:vuln_disc_integral_T}
    S = \int_{T_1}^{T_2} C_{time}t^{-\alpha} = \frac{C_time}{1-\alpha}[T_{2}^{1-\alpha}-T_{1}^{1-\alpha}]
\end{equation}

In one study of black-box mutational fuzzers, the authors probed the software FFmpeg for 18 days and found 400 unique crashes.\cite{bb_mut_fuzz} Most of these crashes are bugs that are not vulnerabilities but we can estimate the fraction that are. Across all the software they tested, they found 363 unique bugs from 4,000 crashes. At that ratio, $C_{time}$ in Eqn \ref{eqn:vuln_disc_integral_T} works out to about 85.5. To compare to humans in the bug bounty program, ten vulnerabilities appears to be a pretty typical count for the first week of the power law. The power law starts in the second week of the program because there is often a deluge of vulnerabilities in the first week. Following Eqn \ref{eqn:vuln_disc_integral_T}, $C_{time}$ is therefore about 6 for the collection of humans.

\begin{figure}[h]
    \label{fig:vuln_disc}
    \centering
    \includegraphics[width=8cm]{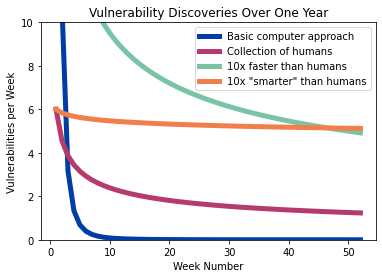}
    \includegraphics[width=8cm]{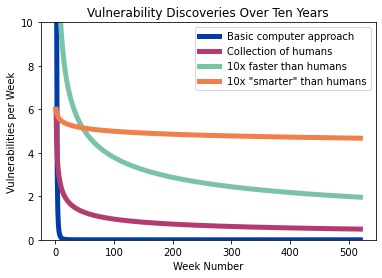}
    \caption{More creative approaches find more vulnerabilities even if they are not as fast. They also continue to find vulnerabilities for a long time.}
\end{figure}

To understand how AI might impact vulnerability discovery, we choose four different scenarios. We use humans in the bug bounty program as an intelligent baseline and the black-box mutational fuzzer as a simple automated baseline. Then we have an AI system that has the same $\alpha$ value as humans for finding new vulnerabilities but operates ten times faster. And finally we include an example where $C$ is kept at the human level but $\alpha$ is ten times smaller to illustrate a system that is better at finding hard-to-find vulnerabilities. Fig \ref{fig:vuln_disc} shows how many new vulnerabilities to expect in a piece of software every week for the first year (a) and the first ten years (b).

One striking observation in Fig \ref{fig:vuln_disc} is that the discovery rate for black-box mutational fuzzers goes to zero after a few months but the other three continue to generate new vulnerabilities even ten years later. The math in Eqn \ref{eqn:vuln_disc_integral_T} explains that if $\alpha$ is greater than one then the tester will saturate after some time but if $\alpha$ is less than one then the tester is creative enough to keep slowly finding vulnerabilities. In the real world the number of vulnerabilities to find cannot actually be infinite so eventually the real discovery rates will slow but that need not be for many years. Even if the number of vulnerabilities is actually finite, humans are not showing signs of those limits yet.

This means that scaling up the computing capacity to try many attempts very quickly (effectively compressing a long search time to a short wall clock duration) is only a good idea while the machine is 'creative' enough to keep finding new vulnerabilities at a high enough rate to continue justify the expense. But it appears that these creative systems will continue to find vulnerabilities rather than exhausting them, so there will simply be more vulnerabilities for defenders to manage and for attackers to exploit.

Whether speed or creativity is more important depends on the timeline being considered. Between a faster and a smarter approach, the faster approach may yield more vulnerabilities initially, but the smarter approach will always eventually churn out vulnerabilities at a higher rate. As long as the smarter approach has $\alpha$ less than 1, it is also guaranteed to produce more total vulnerabilities if it runs long enough. That is beneficial if the goal is to find as many vulnerabilities. But if the goal is to find many vulnerabilities during pre-release or in beta-release so that few are left for attackers, then faster approaches are better than more creative ones. Development of more creative vulnerability discovery mechanisms could be counterproductive by continually flooding the market with vulnerabilities.

\section{Patching vs Exploitation}
\label{sec:patch-exploit}
Once a vulnerability is discovered and disclosed then a race begins where defenders try to develop and distribute a patch faster than attackers can develop and distribute an exploit. For this paper we ignore the deployment time of the exploit because rapid exploit deployment does not require advances in artificial intelligence and because it can already be so fast. In this section we will consider each of the other three phases individually then combine them together. For our analysis, we assume that the timelines in each stage are independent which may not be true in the real world. It may be that vulnerabilities which are faster to exploit are also faster to patch, although patch development timelines do not vary greatly by severity.\cite{patch_dev_timelines}

\subsection{Patch Development}
\label{sec:patch-dev}
The majority of patches are developed before they are ever disclosed. The remaining patches come after varying delays that we model as a Weibull distribution shown in Eqn \ref{eqn:weibull_pdf}. The cumulative Weibull distribution (Eqn \ref{eqn:patch_dev}) therefore estimates the fraction of those vulnerabilities for which a patch is available.

\begin{equation}
    \label{eqn:weibull_pdf}
    f = \frac{k}{\lambda}(\frac{t}{\lambda})^{k-1}e^{-(\frac{t}{\lambda})^{k}}; t>0
\end{equation}
\begin{equation}
\label{eqn:patch_dev}
P_{patch\_developed} = 1 - e^{-(\frac{t}{\lambda})^k}
\end{equation}

\begin{figure}[h]
    \centering
    \includegraphics[width=8cm]{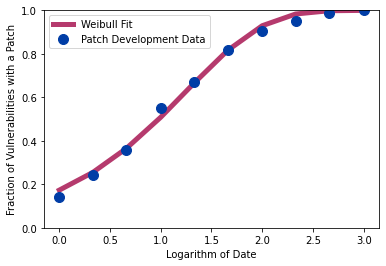}
    \includegraphics[width=8cm]{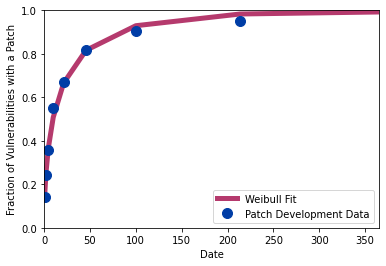}
    \caption{Patch development timelines match well to a Weibull distribution. This includes only the 22\% of patches that are released after the vulnerability is disclosed.}
    \label{fig:patch_dev}
\end{figure}

The Weibull distribution is not as complicated as it looks, it has only two parameters: a shape parameter ($k$) and a scale parameter ($\lambda$). If the shape parameter is one, then it is simply the exponential distribution. That would happen if the developers do not take into account how long a vulnerability has been waiting when deciding whether to write a patch for it. Since the actual shape parameter is less than one, developers appear to be more inclined to write a patch soon after a vulnerability is disclosed than to address ones that have been known for a long time. To find the value more precisely, we fit the Weibull to data we extracted from an empirical study of patch development timelines.\cite{patch_dev_timelines} We estimate the shape parameter to be 0.57 and $\lambda$ to be 18.2 as shown in Fig \ref{fig:patch_dev}.

\begin{figure}[h]
    \centering
    \includegraphics[width=10cm]{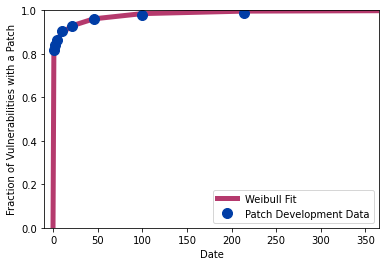}
    \caption{A patch usually exists before a vulnerability is disclosed and a patch is often developed quickly for those that are disclosed first.}
    \label{fig:patch_dev_linear_all_vulns}
\end{figure}

Those patches only include ones that came after the vulnerability was disclosed. We will focus on these vulnerabilities, but the cumulative fraction of vulnerabilities that have a patch available including the 78\% that have patches before they are disclosed is shown in Fig \ref{fig:patch_dev_linear_all_vulns}.

\subsection{Patch Deployment}
\label{sec:patch-dep}
Patch deployment is slower than patch development and is described by a simple exponential distribution (Eqn \ref{eqn:patch_dev}) rather than a Weibull distribution. This exponential distribution has been shown by several researchers, although timelines appear to have shortened somewhat over the last few decades.\cite{patch_adoption_old, patch_adoption_medium} 

\begin{equation}
\label{eqn:patch_dep}
P_{patch\_deployed} = 1 - e^{-\beta t}
\end{equation}

Patch deployment timelines can be different for each vulnerability but also depend on the vulnerable system that is being patched. Some systems update as soon as a patch is available and others may never be able to implement the patch. Empirically, about half of all vulnerabilities have half of their patches adopted within about 100 days.\cite{patch_adoption_2015} That corresponds to a $\beta$ value of 1/144 in Eqn \ref{eqn:patch_dep} which is the value that we  will use as a baseline.

Patch deployment may be somewhat faster than this because the study we drew this data from did not separate their timelines into development and deployment phases. The difference should not be dramatic because most patches are developed prior to disclosure and because most of the remaining ones are developed quickly. Again, for this work we are less concerned with precisely determining the values for our assumptions than understanding the mathematical relationship among them to explore how advances in technology might impact cybersecurity.

\subsection{Overall Patching Timelines}
\label{patching_overall}
Patch deployment can only begin once a patch is developed, so the total delay is the sum of the delay from both development and deployment. Mathematically, the sum of two independent random variables is the convolution of those two probability distributions as shown in Eqn \ref{eqn:eqn_convolution}. 

\begin{equation}
\label{eqn:eqn_convolution}
    P_{patch\_delay}(t) = \int_0^t P_{patch\_developed}(t-\tau)P_{patch\_deployed}(\tau)d\tau
\end{equation}

\begin{figure}[h]
    \centering
    \includegraphics[width=10cm]{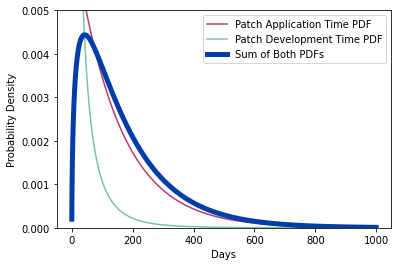}
    \caption{The distribution for the overall patching delay is a convolution between the patch development delay and the deployment delay.}
    \label{fig:patch_convolution}
\end{figure}

We compute the integral in Eqn \ref{eqn:eqn_convolution} numerically as shown in Figure \ref{fig:patch_convolution}.

The combined probability density function has a different shape than either of the two functions in the convolution. The two component functions are both monotonically decreasing but their convolution is not. It rises to a peak then decreases for larger times. That is because although both the Weibull and Exponential have high probability density for small times, most of their probability mass is at higher times. More plainly, although short delays have high probability for each component, there is only a small range of very short delays so very short delays are not all that likely. When sampling from the two distributions then adding them, at least one of the two is likely to contribute a sizeable delay so it is unlikely for the total delay to be very short. As a result, the combined probability distribution starts low and grows with increasing time. It is also unlikely for both of the two component distributions to be very large so the combined probability distribution reaches a peak then decreases.

\subsection{Exploit Development}
\label{sec:exploit-dev}
We only consider the development phase for exploitation. In practice, exploits are not used immediately, but there are few technical barriers to attempting exploitation at scale. Exploit development on the other hand is hard to accelerate. Development timelines have been studied by several authors based on exploits entered in ExploitDB and also based on the internal holdings of professional exploit developers.\cite{exploit_dev_CVSS, ablon_zero_days} 

The ExploitDB study found that the number of vulnerabilities with an exploit was distributed according to an exponentially-capped power law shown in Eqn \ref{eqn:exploit_dev_eqn}.\cite{exploit_dev_CVSS} That is simply a power law multiplied by an exponential. The power law is a growing function and the exponential shrinks so the effect of the exponential is to keep the power law from growing too large, effectively capping it. That cap occurs at a fraction less than one since not all vulnerabilities ever have exploits developed for them.

\begin{equation}
    \label{eqn:exploit_dev_eqn}
    P_{exploit\_development} = At^{a}e^{-bt}
\end{equation}

\begin{figure}[h]
    \centering
    \includegraphics[width=10cm]{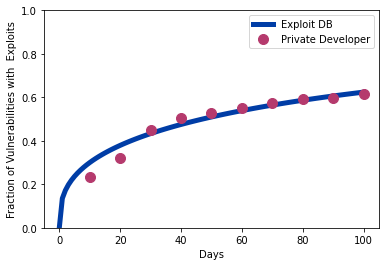}
    \caption{Exploit development delays are fairly consistent whether they are submitted to ExploitDB or created and retained by a private exploit developer.}
    \label{fig:exploit_dev_fig}
\end{figure}

The exponentially-capped power law is shown in Fig \ref{fig:exploit_dev_fig} with the parameters we extracted from the ExploitDB study ($A=0.135$, $a = 0.349$, and $b = 7.90 \times 10^{-4}$). We also compare this distribution to the timelines for exploit development by a private developer.\cite{ablon_zero_days} To do the comparison, we need to account for the vulnerabilities that the developer decided not to exploit. Holding the ExploitDB parameters above fixed, we treat the private developer's normalization constant as a fitting parameter. That total number of vulnerabilities was 239.7 which implies that 79.7 vulnerabilities did not receive exploits given that 160 did. After normalizing to account for unexploited vulnerabilities, Fig \ref{fig:exploit_dev_fig} shows that the two vulnerability timelines are in close agreement. 

\subsection{Patching vs Exploitation Race}
\label{sec:exploit-dev}
\begin{figure}[h]
    \centering
    \includegraphics[width=10cm]{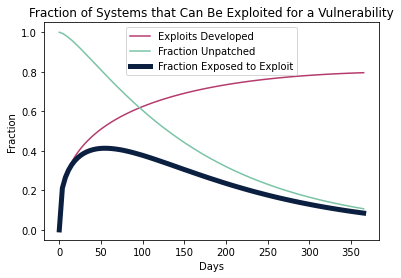}
    \caption{The fraction of systems that are unpatched while there is an existing exploit peaks after a couple months then declines gradually.}
    \label{fig:patch_exploit_explain}
\end{figure}

Having considered each of the phases of the patching vs exploitation race, we are interested in determining the fraction of systems that can be exploited for a given vulnerability. That requires multiplying the probability that a system is unpatched with the fraction of exploits that have been developed as shown in Fig \ref{fig:patch_exploit_explain}. Using the parameters from the empirical studies discussed up to now, the exploitable fraction rises rapidly to a peak of 41\% after 55 days then decays slowly. After a year, 8.5\% of systems are still exploitable.

\begin{figure}[h]
    \centering
    \includegraphics[width=8cm]{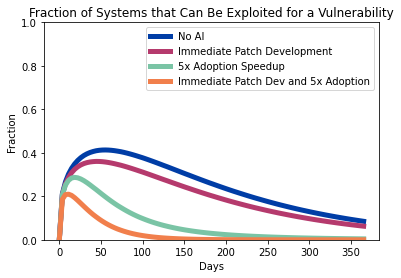}
    \includegraphics[width=8cm]{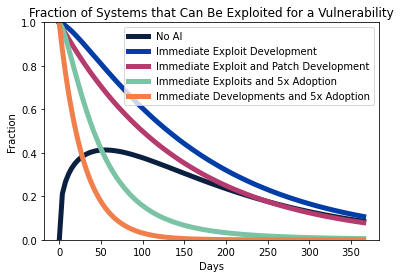}
    \caption{The fraction of exploitable systems does not depend strongly on improvements in patch development but does depend on improvements in exploit development (b) and on improvements in patch deployment (a) and (b).}
    \label{fig:patch_exploit_with_AI}
\end{figure}

 These fractions and timelines can change as technology advances in exploit development, patch development, or patch deployment. Patch development technology is the least impactful of the three. That is because patch development does not provide a substantial delay, so even a technology that completely removed patch development delays would have only a relatively small impact on the exploitable fraction as shown in Fig \ref{fig:patch_exploit_with_AI}a. The total effect would be even less pronounced than the figure shows because this analysis only includes the minority of vulnerabilities that were disclosed before a patch was developed. 
 
 Rapid exploit development on the other hand has a pronounced effect that results in the obvious qualitative differences between part a and part b in Fig \ref{fig:patch_exploit_with_AI}. The curves in part a include delays in exploit development; part b shows the effect of rapid exploit development where one hundred percent of new vulnerabilities would be exploitable. That number would naturally reduce somewhat when considering all vulnerabilities rather than just those that are disclosed before developing a patch but again, patch deployment is the dominant delay, not development. Either way, rapid exploit development would lead to a marked increase from the status quo in a way that patch development technology would not.

Improvements in patch deployment could significantly alleviate that increased exposure but it is a difficult task. Some vulnerable systems cannot be taken offline regularly for updates and many systems or organizations are reluctant to make any alterations without assurances that the updates are stable and will not interfere with workflows. Still, as Fig \ref{fig:patch_exploit_with_AI} shows, partial advances that reduce patch deployment timelines without fully eliminating the delays (i.e. a 5x shorter average patch delay) appear to be more beneficial than technologies for writing patches quickly. Even in a future where exploits are developed instantaneously, partial speedups in patch adoption can lead to drastically shorter exposure periods than the current status quo.

\section{Conclusion}
\label{sec:conclusion}
These simple models only try to capture some highly abstracted behaviors in a few areas of cybersecurity, but they illustrate the potential for quantitatively modeling the impact of technological advances. They suggest that AI may have a more limited impact on phishing than some fear, at least for organizations that are already being targeted. They also suggest a dangerous potential for vulnerability discovery if testing systems are able to find new vulnerabilities for longer rather than just operate faster. And they suggest that there is much more opportunity for automated or accelerated patch deployment than there is for automated patch development.

These models also force us to be specific in terms of their structures and in terms of their inputs. We encourage others to debate those assumptions and improve on them. We hope that increased precision in that debate will help the field coalesce on a set of reasonable bounds to expect of technological advancement and a set of recommended actions or investments that maximize the benefits of technological advances while minimizing the downsides.

\bibliographystyle{unsrt}  
\bibliography{references}

\end{document}